\documentclass[cernpreprint, atlasdraft=false, UKenglish, texmf, orcidlogo]{atlasdoc}
 
\usepackage{atlaspackage}
\usepackage{atlasbiblatex}
 
\usepackage{atlasphysics}

\usepackage{multirow}
 
\usepackage{siunitx}
 
\usepackage{calc}  
\usepackage{array} 
\newlength\mylen
\graphicspath{{logos/}{figures/}}
 
\usepackage{ANA-STDM-2022-17-PAPER-defs}

\AtlasTitle{Measurement of $ZZ$ production cross-sections in the four-lepton final state in $pp$ collisions at $\sqrt{s}=13.6~\TeV$ with the ATLAS experiment}

\AtlasAbstract{
This paper reports cross-section measurements of $ZZ$ production in $pp$ collisions at $\sqrt{s}=13.6~\TeV$ at the Large Hadron Collider. The data were collected by the ATLAS detector in 2022, and correspond to an integrated luminosity of 29~\ifb. Events in the $ZZ\rightarrow4\ell$ ($\ell = e$, $\mu$) final states are selected and used to measure the inclusive and differential cross-sections in a fiducial region defined close to the analysis selections. The inclusive cross-section is further extrapolated to the total phase space with a requirement of 66 $< m_Z <$ 116~\GeV for both $Z$ bosons,
yielding $16.8 \pm 1.1$~pb.
The results are well described by the Standard Model predictions.
}

\AtlasRefCode{STDM-2022-17}
 
\PreprintIdNumber{CERN-EP-2023-235}

\AtlasJournal{Phys. Lett. B}
\AtlasJournalRef{\PLB 855 (2024) 138764}
\AtlasDOI{10.1016/j.physletb.2024.138764}

\hypersetup{pdftitle={ATLAS document},pdfauthor={The ATLAS Collaboration}}
 
\begin{document}
 
\maketitle

\section{Introduction}
\label{sec:intro}

% The next lines are included from the .//sections/introduction.tex input file
The study of the production of pairs of massive electroweak gauge bosons (`dibosons') has an important place in the physics programme at the Large Hadron Collider (LHC). Among all the contributing Feynman diagrams, those containing triple-gauge-boson couplings (TGCs) or involving the Higgs boson are particularly interesting, due to their direct connections with two fundamental features of the Standard Model (SM) electroweak (EW) theory, namely the non-Abelian gauge group structure and spontaneous EW symmetry breaking. Moreover, diboson production is a key process used to directly search for new phenomena beyond the SM (BSM). The diboson studies can be traced back to experiments at LEP~\cite{ALEPH:2013dgf} and Tevatron~\cite{D0:2013rca, CDF:2014nef}; the LHC extends the precision measurements to unprecedented energy scales,
owing to the high centre-of-mass energy and the large integrated luminosity.
 
The production of two on-shell $Z$ bosons is the rarest diboson process, but still attractive
because of the high signal-to-background ratio for the fully leptonic decay channels. The $ZZ$ process is a key channel to search for anomalous neutral TGCs (aTGCs)~\cite{Baur:2000ae} and to study off-shell production of the Higgs boson~\cite{Goncalves:2017iub}. Various measurements of $ZZ$ production have been performed by the ATLAS and CMS collaborations using data taken at $\sqrt{s} = $ 5.02, 7, 8, and 13~\TeV~\cite{PhysRevLett.127.191801, ATLAS:2012bra,ATLAS:2016bxw, CMS-SMP-12-007, CMS-SMP-12-016, CMS-SMP-13-005, CMS-SMP-19-001, CMS-SMP-16-001, CMS-SMP-19-001, ATLAS:2017bcd, ATLAS:2019xhj}. Two complementary final states originating from the decays of the $Z$ boson pair are utilised in those measurements, one with four charged leptons (\zzfourl), and the other with two charged leptons and two neutrinos (\zztwoltwov), where $\ell$ refers to $e$ or $\mu$. The smallest relative error in the measured inclusive cross-section was achieved with 13 \TeV~data, and is 5\%~\cite{ATLAS:2017bcd} and 7\%~\cite{ATLAS:2019xhj} for the \zzfourl~and \zztwoltwov~channels, respectively. Differential measurements have been performed on kinematic variables sensitive to
aTGCs and off-shell Higgs boson production.
These measurements also cast light on calculations of higher-order quantum chromodynamics (QCD) and EW corrections.
Two key variables commonly considered for the \zzfourl~channels are the invariant mass and the transverse momentum of the four-lepton system, \mfourl~ and \ptfourl, respectively, that are studied here.
 
This paper reports a first measurement of $ZZ$ production in $pp$ collisions at $\sqrt{s}=$ 13.6~\TeV, which extends the
test of the SM for this rare process to
a new $pp$ centre-of-mass energy.
The \zzfourl~channel, where $\ell$ can be an electron or a muon, is utilised in this measurement, and a signal region is defined by selecting four high-\pt, isolated leptons, with a fiducial region defined close to the detector-level selections. The cross-section for \zzfourl~production is measured in the fiducial region, and extrapolated to the total $ZZ$~production cross-section satisfying 66 $< m_Z <$ 116~\GeV. Finally, the differential measurements are presented.

% End of text imported from the .//sections/introduction.tex input file

\section{The ATLAS detector}
\label{sec:detector}

% The next lines are included from the .//sections/detector.tex input file
\newcommand{\AtlasCoordFootnote}{
ATLAS uses a right-handed coordinate system with its origin at the nominal interaction point (IP)
in the centre of the detector and the \(z\)-axis along the beam pipe.
The \(x\)-axis points from the IP to the centre of the LHC ring,
and the \(y\)-axis points upwards.
Polar coordinates \((r,\phi)\) are used in the transverse plane,
\(\phi\) being the azimuthal angle around the \(z\)-axis.
The pseudorapidity is defined in terms of the polar angle \(\theta\) as \(\eta = -\ln \tan(\theta/2)\).
Angular distance is measured in units of \(\Delta R \equiv \sqrt{(\Delta\eta)^{2} + (\Delta\phi)^{2}}\).}
 
The ATLAS detector~\cite{PERF-2007-01,ATLAS:2023dns} at the LHC is a multipurpose particle detector
with a forward--backward symmetric cylindrical geometry and a near \(4\pi\) coverage in
solid angle.\footnote{\AtlasCoordFootnote}
It consists of an inner tracking detector surrounded by a thin superconducting solenoid
providing a \qty{2}{\tesla} axial magnetic field, electromagnetic and hadron calorimeters, and a muon spectrometer.
The inner tracking detector covers the pseudorapidity range of \(|\eta| < 2.5\).
It consists of silicon pixel, silicon microstrip, and transition radiation tracking detectors.
Lead/liquid-argon (LAr) sampling calorimeters provide electromagnetic (EM) energy measurements
with high granularity.
A steel/scintillator-tile hadron calorimeter covers the central pseudorapidity range of \(|\eta| < 1.7\).
The endcap and forward regions are instrumented with LAr calorimeters
for both the EM and hadronic energy measurements up to \(|\eta| = 4.9\).
The muon spectrometer surrounds the calorimeters and is based on
three large superconducting air-core toroidal magnets with eight coils each.
The field integral of the toroids ranges between \num{2.0} and \qty{6.0}{\tesla\metre}
across most of the detector.
The muon spectrometer includes a system of precision tracking chambers and fast detectors for triggering.
A two-level trigger system is used to select events.
The first-level trigger is implemented in hardware and uses a subset of the detector information
to accept events at a rate below \qty{100}{\kHz}.
This is followed by a software-based trigger that
reduces the accepted event rate to \qty{3}{\kHz} on average,
depending on the data-taking conditions.
An extensive software suite~\cite{ATL-SOFT-PUB-2021-001} is used in data simulation, in the reconstruction
and analysis of real and simulated data, in detector operations, and in the trigger and data acquisition
systems of the experiment.

% End of text imported from the .//sections/detector.tex input file

\section{Data and simulation}
\label{sec:data_mc}

% The next lines are included from the .//sections/samples.tex input file
 
This measurement uses the $pp$ collision data recorded by the ATLAS detector in 2022, the first year of the Run~3 data taking period. The data sample corresponds to an integrated luminosity of $29$~\ifb, with an uncertainty of 2.2\%~\cite{ATL-DAPR-PUB-2023-001}.
The mean number of interactions per bunch crossing, known as pile-up, averaged over all colliding bunch pairs is about 40.
Events are required to have fired at least one single-lepton trigger~\cite{TRIG-2018-01,TRIG-2018-05}.
For the main triggers, the transverse momentum (\pt) thresholds are 26 and 24~\GeV\ for electrons and muons, respectively. The combined trigger efficiency is above $99\%$ for events satisfying the signal selections given in Section~\ref{sec:select}.
 
Simulated data samples from Monte Carlo (MC) generators are used to estimate detector effects,
and to provide theoretical predictions
for comparison with the measured results.
The MC samples employed are summarised in the following.
The next-to-next-to-leading-order (NNLO) parton distribution function (PDF) set of \NNPDF[3.0]~\cite{Ball:2014uwa} was used in the matrix-element (ME) calculation for all processes.
 
\begin{figure}[!htbp]
\centering
\subfloat[]{\includegraphics[width = 0.24\textwidth]{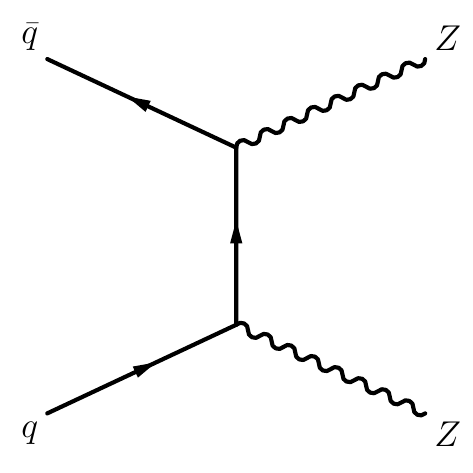}}
\subfloat[]{\includegraphics[width = 0.24\textwidth]{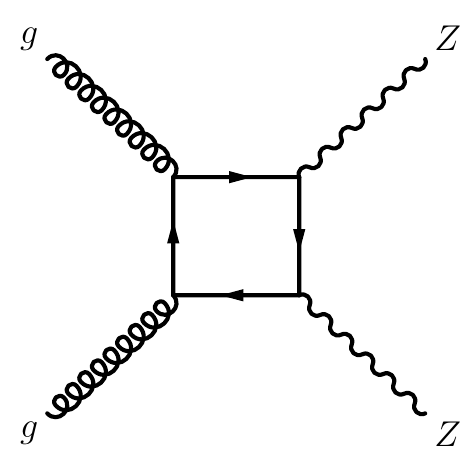}}
\subfloat[]{\includegraphics[width = 0.24\textwidth]{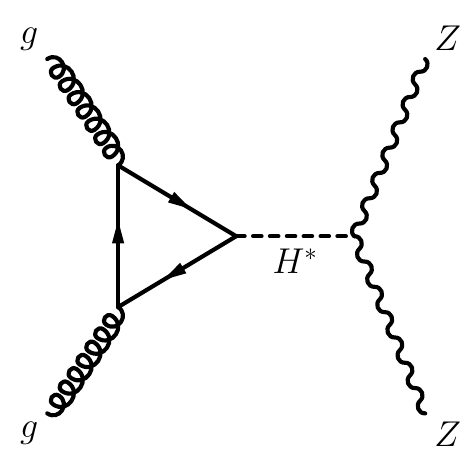}}
\subfloat[]{\includegraphics[width = 0.24\textwidth]{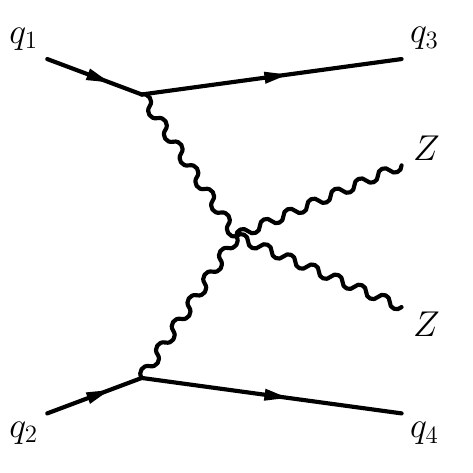}}\\
\caption{Examples of leading-order SM Feynman diagrams for different classes of $ZZ$ production in proton–proton collisions: (a)
\qqZZ, (b) \ggZZ, (c) \ggtoHtoZZ\ and (d) electroweak \qqZZjj.
\label{fig:diagrams}}
\end{figure}
 
The dominant contribution to $ZZ$ production, the $q\bar{q}$-initiated process (\qqZZ), was generated with \progname{Sherpa}~2.2.12~\cite{Gleisberg:2008ta,Gleisberg:2008fv,Cascioli:2011va}, with ME calculation accuracy at next-to-leading order (NLO) in the strong coupling $\alpha_s$ for final states with up to one additional parton, and at leading-order (LO) accuracy for up to three additional partons.
The ME calculations utilised \progname{OpenLoops}~\cite{Cascioli:2011va} for the virtual QCD corrections, and
were merged with the \progname{Sherpa} parton shower (PS) using the ME+PS@NLO prescription~\cite{Hoeche:2012yf}.
The \qqZZ~contribution also includes $\ensuremath{qg\rightarrow ZZq}$ production processes.
\progname{Sherpa}~2.2.12 provides higher-order NLO EW corrections as alternative internal weights~\cite{Biedermann:2017yoi}.
The predicted cross-section in the fiducial region for this measurement, defined in Section~\ref{sec:select}, is reduced by 12\% when adopting the alternative weights, which then lies below other theoretical predictions~\cite{Biedermann:2016lvg, Grazzini:2015hta} by approximately 10\%. Therefore the NLO EW weights are not used to estimate detector effects, and are only shown when comparing the measured data with predictions.
 
The loop-induced gluon-initiated signal (\ggZZ) was simulated with \progname{Sherpa}~2.2.12 at LO precision with up to one extra parton, with the same PS modelling as described above.
The \ggZZ~sample includes the box-diagram \ggZZ~process and the off-shell Higgs production with $H^{*} \rightarrow ZZ$ decays.
The predicted cross-section of \ggZZ~is scaled by a $K$-factor of 1.7 to account approximately for the higher-order QCD corrections. The $K$-factor was calculated for massless quark loops~\cite{PhysRevD.92.094028,Campbell_2016} in the heavy top-quark approximation~\cite{MELNIKOV201543}. A relative uncertainty of 60\% was assessed from QCD scale variations~\cite{ATLAS:2020tlo}.
 
EW production of $ZZ$ in association with two jets (EW \qqZZjj) was modelled with \POWHEGBOX[v2]\cite{Hamilton:2013fea,Hamilton:2015nsa,Alioli:2010xd,Nason:2004rx,Frixione:2007vw,Jager:2013iza} at LO in the EW coupling, corresponding to $\alpha^{4}$, and at NLO in $\alpha_s$. The generator was interfaced to \PYTHIA[8.3]~\cite{Bierlich2022} for PS and hadronisation.
 
These three sets of processes with the subsequent \zzfourl~decays provide the signal in this measurement. Figure~\ref{fig:diagrams} shows representative Feynman diagrams for the respective contributions to $ZZ$ production.
 
Additional samples were generated to estimate the background contribution in the selected phase space.
Irreducible backgrounds arise from $t\bar{t}Z$ and tri-boson ($VVV$) production, where $V = W, Z$, both modelled with \progname{Sherpa}~2.2.12, requiring at least four charged leptons in the final state.
Backgrounds with non-prompt leptons (Section~\ref{sec:background}) arise primarily from $Z$ boson production with associated jets ($Z+$jets),
production of $WZ$ and $t\bar{t}$. Production of $Z+$jets was simulated using \progname{Sherpa}~2.2.12 with up to two partons at NLO and four partons at LO in $\alpha_s$.
The $WZ$ process was simulated using \progname{Sherpa}~2.2.12, with a similar configuration as for the \qqZZ~signal.
Events from $t\bar{t}$ production were generated with \POWHEGBOX[v2]+\PYTHIA[8.3] at NLO in $\alpha_s$.
 
A full simulation of detector effects using \GEANT~\cite{Agostinelli:2002hh,SOFT-2010-01} was applied to all these samples, along with dedicated calibrations and corrections to physics objects to match the performance of real data.

% End of text imported from the .//sections/samples.tex input file

\section{Object and phase space definitions}
\label{sec:select}

% The next lines are included from the .//sections/selection.tex input file
 
To account for the performance of the ATLAS detector and the \zzfourl~final-state signature, the candidate events must satisfy requirements on data-taking quality, triggers (Section~\ref{sec:data_mc}), physics objects and reconstructed boson candidates, to improve the signal purity by suppressing the backgrounds.
The data quality requirements are imposed to reject running periods in which detector subsystems were not operating robustly. All candidate events have at least one primary vertex~\cite{ATL-PHYS-PUB-2019-015} with at least two associated inner tracking detector tracks.
The object and event selections, which are described below, are similar to those defined for previous ATLAS measurements~\cite{ATLAS:2017bcd,ATLAS:2021kog}.
 
Electrons are reconstructed from energy deposited in the electromagnetic calorimeter matched to tracks in the inner tracking detector.
Electrons are required to have $\pt>7$~\GeV and $|\eta|<2.47$. Candidates within the transition region between barrel and endcap calorimeters ($1.37 < |\eta| < 1.52$) are excluded.
In addition, to distinguish real, isolated electrons from other objects, `LooseAndBLayer' likelihood identification and 
`Loose' isolation selections~\cite{EGAM-2018-01,ATLAS:2019jvq}, both optimised for Run~3 analysis conditions, must be satisfied.
 
Muons are reconstructed~\cite{PERF-2015-10} primarily by matching the tracks in the muon spectrometer and inner tracker.
Because of the limited geometrical coverage of the muon spectrometer at the centre of the detector ($|\eta| < 0.1$), a fully reconstructed inner detector track is also matched to a muon spectrometer track segment or a calorimeter energy deposition (calorimeter-tagged muons) for muon identification.
Muon candidates are required to have $\pt
> 5~\gev$ and $|\eta| < 2.5$.
The $\pt$ threshold is increased to $15$~\GeV for calorimeter-tagged muons.
Candidates are required to further satisfy the `Loose' identification and `PflowLoose' isolation~\cite{PERF-2015-10,ATLAS:2020auj} quality requirements.
 
To select lepton candidates originating from the primary vertex, a requirement is placed on the longitudinal impact parameter, $z_0$, multiplied by the sine of the track polar angle $\theta$ with $|z_0 \times \sin\theta| < 0.5$ mm. A requirement is also placed on the significance of the transverse impact parameter $d_0$ with $|d_0|/\sigma_{d_0}<5 (3)$ for electrons (muons).
 
Events with at least two individual pairs of same-flavour-opposite-charge (SFOC) leptons are selected. The two highest-$\pt$ leptons are required to have $\pt>$ 27~\GeV and 10~\GeV, respectively.
Any SFOC pair must have an invariant mass greater than 5~\GeV, removing low-mass resonance contributions. Any lepton in the final state must be separated from other leptons with the requirement of $\Delta R(\ell_{i},\ell_{j}) > 0.05$.
In each event with at least four leptons, the quadruplet of leptons is formed from two separate SFOC pairs.
The SFOC pair with mass closest to the $Z$ mass~\cite{ALEPH:2005ab} is called the leading pair ($\ell\ell,1$). Then the subleading pair ($\ell\ell,2$) is formed from the remaining leptons, with mass next closest to the $Z$ mass.
In addition, the two SFOC lepton pairs are required to satisfy $66 < m_{\ell\ell,1}, m_{\ell\ell,2} < 116$~\GeV. The invariant mass of the four leptons in the two pairs is required to be $m_{4\ell} > 180$~\GeV, motivated to select only on-shell $ZZ$ events.
Candidate events selected with these criteria enter into the signal region (SR) for this measurement.

% End of text imported from the .//sections/selection.tex input file

% The next lines are included from the .//sections/fiducialregion.tex input file
The fiducial phase space is defined with a set of criteria very close to that of the detector-level SR criteria. This strategy helps to reduce the amount of phase-space extrapolation in the fiducial measurements and therefore reduces the theoretical uncertainties of the results. The criteria are applied to `particle-level' physics objects, which are reconstructed from stable final-state particles, before their interactions with the detector.
For electrons, QED final-state radiation is partly recovered by adding to the lepton four-momentum the four-momenta of surrounding photons not originating from hadrons within an angular distance $\Delta R < 0.1$ (dressed leptons), while the bare leptons defined after QED radiation are considered for muons. The different choices are to best mimic the corresponding object reconstruction at the detector level.
 
The total lepton phase space is defined for both individual SFOC pairs having $66< m_{\ell\ell} <116$~\GeV, and any SFOC lepton pair in the decay final state must satisfy the requirement of $m_{ij} > 5$~\GeV, where $m_Z$ and $m_{ij}$ are calculated with the leptons defined at Born level, i.e. before any QED final-state radiation. Table~\ref{tab:FiducialSelection} summarises the total and fiducial phase-space definitions.
The cross-section in the total phase space is calculated by correcting for the branching fraction for \zzfourl~decays, thus maintaining the invariant mass selections on $Z$ and $ZZ$ systems.

\begin{table}[ht]
\centering
\caption{Definition of the fiducial and total lepton phase-space regions.\label{tab:FiducialSelection}}
\begin{tabular}{ l |  c   c}
\hline
&\textbf{Fiducial phase space} & \textbf{Total lepton phase space} \\
\hline
Muon selection &  Bare, $\pt > 5$~\GeV$, |\eta| < 2.5$ & Born \\
Electron selection &  Dressed, $\pt > 7$~\GeV$, |\eta| < 2.47$ & Born\\
\hline
Four-lepton signature & $\geq$~2 SFOC pairs & $\geq$~2 SFOC pairs \\
Lepton kinematics   &   $\pt > 27 / 10$~\GeV{}  &   \\
Lepton separation               &    $\Delta R(\ell_{i},\ell_{j}) > 0.05$  &   \\
Low-mass $\ell^{+}\ell^{-}$ veto &    $  m_{ij} > 5$~\GeV  &  $m_{ij} > 5$~\GeV  \\
$Z$ mass window   &$66 < m_{\ell\ell,1}, m_{\ell\ell,2} < 116$~\GeV &  $66 < m_{\ell\ell,1}, m_{\ell\ell,2} < 116$~\GeV \\
$ZZ$ on-shell   & $m_{4l} > 180~\GeV$ &  \\
\hline
\end{tabular}
\end{table}

% End of text imported from the .//sections/fiducialregion.tex input file

\section{Background estimation}
\label{sec:background}

% The next lines are included from the .//sections/background.tex input file
 
The background contamination in the SR is expected to be less than 5\% of the total yield with the majority being from reducible backgrounds. The irreducible backgrounds consist almost entirely of $t\bar{t}Z$~ and triboson production. They are evaluated using MC simulation, as discussed in Section~\ref{sec:data_mc}. The uncertainties in these backgrounds, experimental and theoretical, are discussed in Section~\ref{sec:syst}. The final states with $\tau$-leptons from $Z$ decays contribute at the per-mille level and are neglected.
 
A prompt lepton refers to a final-state lepton that does not originate from the decay of a hadron or $\tau$-lepton. Events with two or three prompt leptons can satisfy the SR criteria, if associated non-prompt leptons, jets or photons are mis-identified as signal leptons (fake leptons), leading to a reducible fake background. Most reducible background arises from fake leptons in $Z+$jets and $t\bar{t}$ events. Fake contributions from events with fewer than two prompt leptons are negligible and neglected.
 
Modelling of such background using purely MC predictions is unreliable and therefore a 
fake-factor (FF) technique that estimates the background with real data is used.
The FF is calculated in a region dominated by single $Z$ production, with events that always have a selected SFOC lepton pair, with an invariant mass satisfying $66 < m_{\ell\ell} < 116$~\GeV. The signal trigger and data-quality selections are applied.
Any extra lepton in this region is considered to be fake after subtracting other prompt contributions from $WZ$ and $ZZ$ production estimated from simulation.
The FF is then calculated as the ratio of the number of leptons satisfying the standard lepton selection to the number satisfying the same selection except failing to meet the isolation or transverse impact parameter requirements (fake-enriched selection). A `Loose' likelihood identification~\cite{ATLAS:2019jvq} is additionally required in the fake-enriched selection for electrons, to suppress the overwhelming photon contributions.
The fake factors are obtained in bins of \pt~and $\eta$, separately for electrons and muons.
 
An application region is defined using all the SR criteria, except that one or two leptons are required to satisfy the fake-enriched selection instead of the standard lepton selection. The number of reducible background events in the SR is calculated by applying the FFs to fake-enriched 
events in the application region, after subtracting the contribution from $ZZ$ production, following the methodology described in Ref.~\cite{ATLAS:2017bcd}.
The inclusive yield is reported in Table~\ref{tab:eventyields} and the relative uncertainty is assessed to be 30\%, as discussed in Section~\ref{sec:syst}.
 
The shape of the reducible background is obtained by applying the FF method in each kinematic bin and it is subject to large statistical uncertainties due to the limited sample size in data in the application region. To reduce these statistical fluctuations, a smoothing procedure
based on a variable-span smoothing technique~\cite{Friedman:155846} is utilised.
The smoothed fake distribution and uncertainty are estimated with 50~000 MC sets generated using random values taken from a Gaussian distribution for each bin before smoothing.
Then the smoothing is performed on all the sets. The mean value and standard deviation of the sets in each bin are taken as the nominal value and uncertainty of the smoothed fake background.
In addition, to evaluate the uncertainty from the choice of smoothing technique, Lowess smoothing~\cite{Lowess} is used as an alternative. The smoothing procedure significantly reduces the bin-by-bin fluctuations and improves the background predictions for the differential cross-section measurements.

% End of text imported from the .//sections/background.tex input file

\section{Systematic uncertainties}
\label{sec:syst}

% The next lines are included from the .//sections/systematics.tex input file
The systematic uncertainties can be categorised into experimental and theoretical uncertainties. The uncertainties in the overall cross-section measurements from the sources listed in Table~\ref{tab:fidXsUncer} are given there.
 
Experimental uncertainties address the imprecision in luminosity determination and modelling of detector effects. In addition, a MC statistical uncertainty arises from the limited size of the simulated samples.
The integrated luminosity impacts both the signal and background expected yields.
The uncertainty is 2.2\%~\cite{ATL-DAPR-PUB-2023-001}.
The pile-up modelling uncertainty is assessed by varying by $\pm 4\%$
the value of the visible inelastic cross-section that reweights the pile-up distribution in the simulation to that observed in the data~\cite{STDM-2015-05}.
 
Uncertainties in the lepton-energy calibration, and in the efficiency of lepton reconstruction, identification, isolation and trigger, influence the expected yields and distributions, and also the migration matrices and acceptance and correction factors in the cross-section measurements, defined below.
The estimates are based on studies of the detector performance using Run~2 and Run~3 data~\cite{EGAM-2018-01,PERF-2017-03,PERF-2015-10}.
 
Several sources of uncertainty affect the estimate of the reducible backgrounds, coming from the limited number of data events in the application region, the uncertainty in the subtracted prompt-lepton contributions in that region, and additional uncertainties in the FF estimates~\cite{ATLAS:2017bcd}.
The contribution from the production of $VVV$~and $t\bar{t}Z$~processes to the SR is relatively small, and constant relative uncertainties of 10\% and 15\% are assigned to them, respectively, to account for the total systematic uncertainty covering experimental and theoretical sources.
 
The theoretical uncertainties in modelling the \qqZZ~process are considered for sources including PDF sets, 
higher-order QCD corrections and parton shower models.
The PDF uncertainty is estimated by using 100 eigenvector variations of the \NNPDF[3.0] set, as suggested in Ref.~\cite{Ball:2014uwa}. In addition, the nominal prediction is compared with another prediction employing an alternative PDF set, the NLO \CT[18]~\cite{Hou:2019efy}, to evaluate the uncertainty due to the choice of PDF set. The higher-order QCD uncertainties are evaluated by varying the default choice of renormalisation and factorisation scales independently by factors of two, removing combinations where the variations differ by a factor of four. The largest deviations are taken as the uncertainty.
The impact of parton shower uncertainties are calculated following the recommendation of the \progname{Sherpa}~\cite{Hoeche:2012yf} authors. 
 
For the \ggZZ~process, a conservative QCD correction uncertainty of 60\% is assigned~\cite{PhysRevD.92.094028,ATLAS:2020tlo}, thus the other sources of theoretical uncertainties are not added, as they are assumed to be small compared to the QCD uncertainty. For the EW \qqZZjj~ process, a constant theoretical uncertainty of 20\% is assigned~\cite{STDM-2017-19}.

% End of text imported from the .//sections/systematics.tex input file

\section{Results}

% The next lines are included from the .//sections/measurement.tex input file
\begin{figure}[!htbp]
\centering
\subfloat[]{\includegraphics[width = 0.45\textwidth]{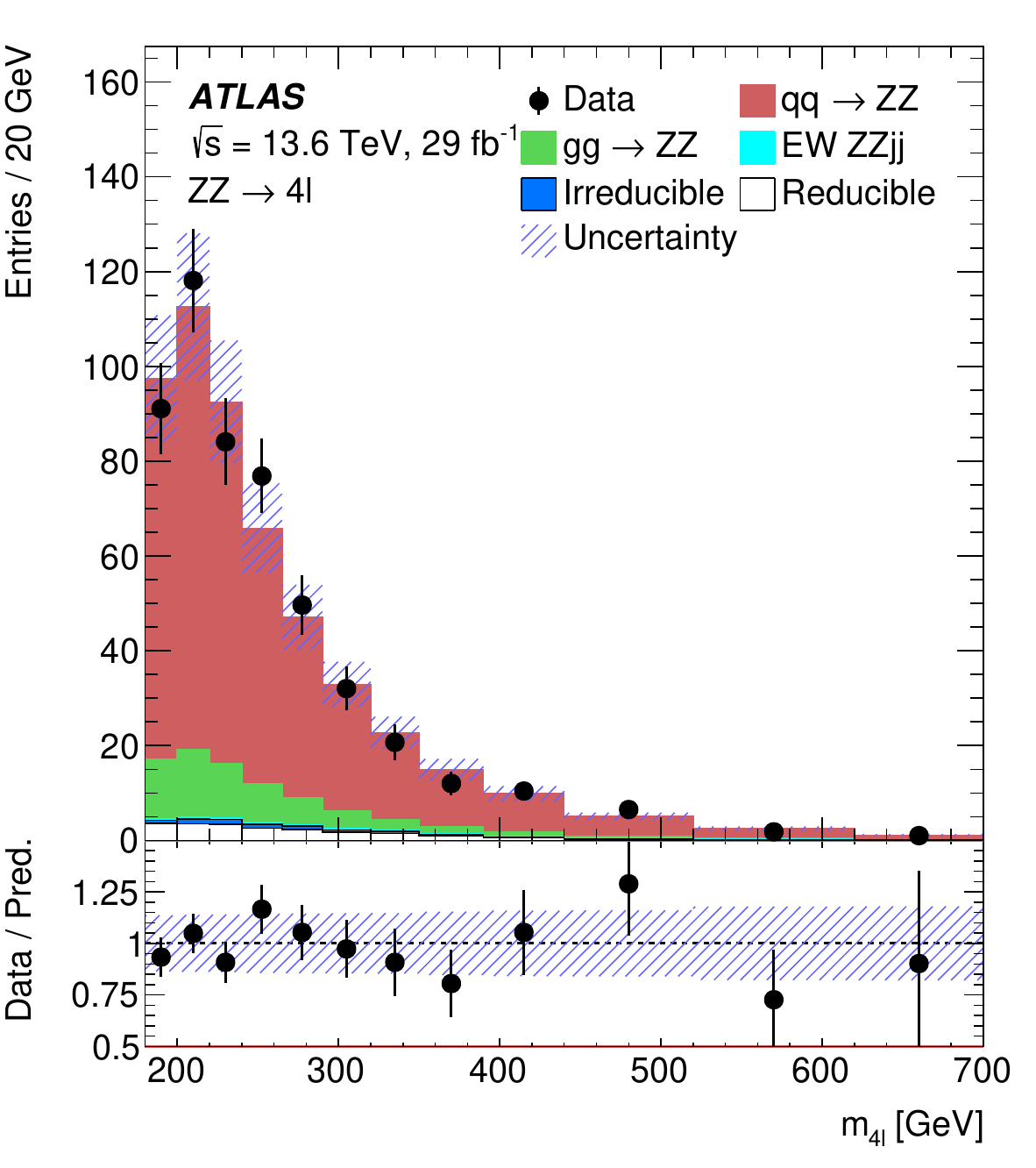}}
\subfloat[]{\includegraphics[width = 0.45\textwidth]{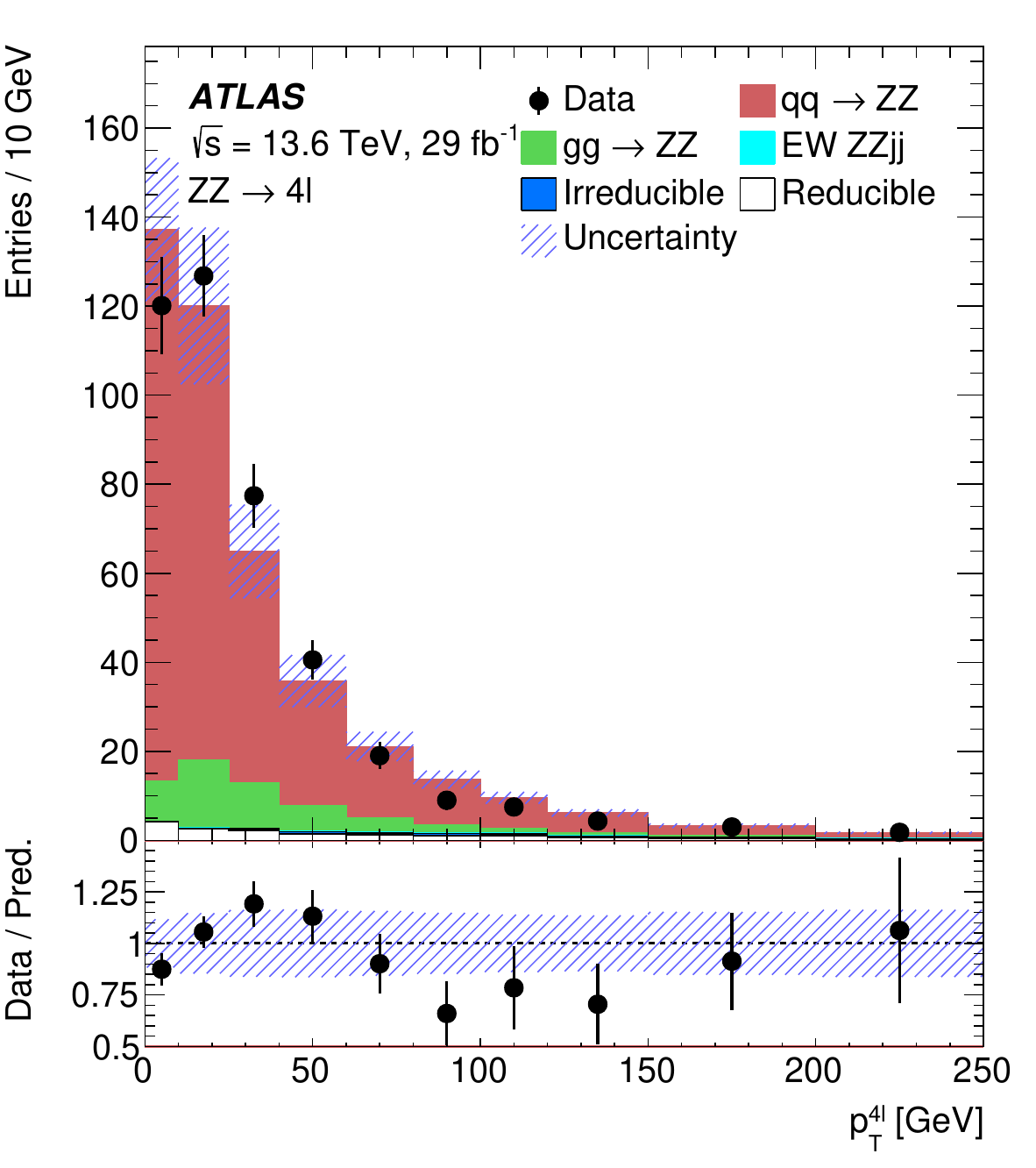}}\\
\caption{Kinematic distributions for (a) \mfourl~and (b) \ptfourl~in the signal region. Data are compared with the predictions with all uncertainties included. The lower panels show the ratios of the data to the predictions.
\label{fig:datavsmc1}}
\end{figure}
 
The observed data yield and the estimated signal and background contributions in the SR are listed in Table~\ref{tab:eventyields}, while Figure~\ref{fig:datavsmc1} gives the detector-level distributions for \mfourl~and \ptfourl.
Good agreement is found  between the observations and predictions. Using these samples, the measurements of the inclusive and differential cross-sections are presented in this section.
 
\begin{table}[ht]
\centering
\caption{
Observed and predicted detector level yields in the signal region, corresponding to the integrated luminosity of 29 \ifb.
All systematic uncertainties described in Section~\ref{sec:syst} are included.}
\label{tab:eventyields}
\resizebox{\textwidth}{!}{
\begin{tabular}{c | c c c c c c | c | c}
\hline
Process & \qqZZ & \ggZZ & EW \qqZZjj & $t\bar{t}Z$ & $VVV$ & Reducible & Total & Data \\
\hline
Yield & 515 $\pm$ 50 & 74 $\pm$ 44 & 4.7 $\pm$ 1.0 & 5.5 $\pm$ 0.8 & 2.1 $\pm$ 0.2 & 25.4 $\pm$ 8.1 & 626 $\pm$ 88 & 625 \\
\hline
\end{tabular}}
\end{table}

% End of text imported from the .//sections/measurement.tex input file

\subsection{Measurement of inclusive cross-sections}
\label{sec:inclXsec}

% The next lines are included from the .//sections/inclXsection.tex input file
The inclusive cross-section in the fiducial region is calculated as:
\begin{equation*}
\sigma_{\mathrm{fid}} = \frac{N_{\mathrm{obs}}-N_{\mathrm{bkg}}}{L\times C_{ZZ}},
\end{equation*}
where $N_{\mathrm{obs}}$ and $N_{\mathrm{bkg}}$ denote the data yield and the expected background contribution in the SR, respectively. The integrated luminosity of 29 \ifb~is denoted by $L$, while $C_{ZZ}$ is the correction factor to account for reconstruction inefficiencies and resolution effects.
 
The $C_{ZZ}$ factor is computed as the ratio of the number of simulated detector-level signal events passing the SR criteria to the number of particle-level events satisfying the fiducial selection. The value of $C_{ZZ}$ is determined to be $0.555 \pm 0.022$, where the uncertainty includes all sources discussed, and is dominated by the uncertainties from electron identification and muon isolation. The theoretical uncertainties have little impact on $C_{ZZ}$ compared to that on the overall normalisation, as they cancel out in the ratio. The fiducial cross-section is measured to be:
\begin{equation} \label{eq:fidxs_result}
\sigma_{\mathrm{fid}} = 36.7 \pm 1.6 (\mathrm{stat}) \pm 1.5 (\mathrm{syst}) \pm 0.8 (\mathrm{lumi}) ~\mathrm{fb}. \notag
\end{equation}
The measurement precision is 6.3\%, with a breakdown into categories shown in Table~\ref{tab:fidXsUncer}. The statistical uncertainty of the data is at the same level as the total systematic uncertainty.
As described in Section~\ref{sec:syst}, the theoretical uncertainty consists of uncertainties from the PDF sets, QCD corrections and parton showering of the signal processes. The uncertainty due to the background estimate contains all sources of uncertainties in the evaluation of reducible and irreducible backgrounds.
 
\begin{table}[ht]
\centering
\caption{Breakdown of relative uncertainty in the measured fiducial cross-section $\sigma_{\mathrm{fid}}$.\label{tab:fidXsUncer}}
\begin{tabular}{ l c }
\hline
Source &  Relative uncertainty(\%) \\
\hline
Data statistical uncertainty  & 4.2 \\[0.1cm]
MC statistical uncertainty & 0.3 \\[0.1cm]
Luminosity  & 2.2  \\[0.1cm]
Pile-up  & 0.3  \\[0.1cm]
Lepton momentum  & 0.2 \\[0.1cm]
Lepton efficiency  & 3.7 \\[0.1cm]
Background & 1.6 \\[0.1cm]
Theoretical uncertainty & 1.0 \\ [0.1cm]
\hline
Total & 6.3 \\[0.1cm]
\hline
\end{tabular}
\end{table}
 
The fiducial cross-section is extrapolated to the total phase space, by dividing it by the acceptance factor ($A_{ZZ}$) and the branching fraction for \zzfourl~decays with $\ell=e,\mu$. The $A_{ZZ}$ factor is the ratio of the cross-section in the fiducial region to that in the total lepton phase space, calculated from theoretical predictions at particle level.
In this measurement, the $A_{ZZ}$ factor is determined by the MATRIX programme~\cite{Grazzini:2015hta}.
MATRIX predicts the cross-section for \zzfourl~processes at fixed order, with next-to-next-to-leading-order QCD and NLO EW accuracy for the \qqZZ~process, and with NLO QCD accuracy for the \ggZZ~process. The \NNPDF[3.0]~PDF set is utilised in the MATRIX calculation.
The relative PDF uncertainty from \progname{Sherpa} is applied to the MATRIX prediction because MATRIX does not support the calculation of PDF variations.
The value of $A_{ZZ}$ is $0.482 \pm 0.003$,
with the uncertainty coming from the PDF set and the QCD scale.
The value of $A_{ZZ}$ is alternatively derived from the signal samples; a good agreement is found with that from MATRIX\@. The total cross-section is:
\begin{equation} \label{eq:totalxs_result}
\sigma_{\mathrm{total}} = 16.8 \pm 0.7 (\mathrm{stat})\pm 0.7 (\mathrm{syst}) \pm 0.4 (\mathrm{lumi}) ~\mathrm{pb}. \notag
\end{equation}
The relative uncertainty is 6.5\%.
 
The measured inclusive cross-sections in the fiducial and total phase spaces are presented in Table~\ref{tab:incluXs}, and compared with predictions. The prediction from MC simulation includes \qqZZ, \ggZZ, and EW \qqZZjj~processes, with the accuracy described in Section~\ref{sec:data_mc}.
The uncertainty in the predicted cross-section from MC simulation includes the theoretical uncertainty sources described in Section~\ref{sec:syst}, and is dominated by the QCD-correction error from the $K$-factor for \ggZZ. The MATRIX prediction includes the QCD scale variation, and the PDF uncertainty from \progname{Sherpa}.
The measured cross-sections are well described by the predictions.

\begin{table}[ht]
\renewcommand\arraystretch{1.5}
\centering
\caption{The measured inclusive cross-sections compared with the predictions from MC simulation and from the MATRIX calculation at fixed order. Both the fiducial and total phase space results are included. In all predictions, EW \qqZZjj~ is modelled by \POWHEGBOX[v2]. \label{tab:incluXs}}
\begin{tabular}{c| c |c |c}
\hline
& Measurement & MC prediction & MATRIX + EW $ZZjj$\\
\hline
Fiducial & $36.7 \pm 1.6 (\mathrm{stat}) \pm 1.5 (\mathrm{syst}) \pm 0.8 (\mathrm{lumi})$ fb & $36.8~^{+4.3}_{-3.5}$ fb & $36.5 \pm 0.7$ fb \\[0.1cm]
\hline
Total & $16.8 \pm 0.7 (\mathrm{stat})\pm 0.7 (\mathrm{syst}) \pm 0.4 (\mathrm{lumi})$ pb & $17.0~_{-1.4}^{+1.9}$ pb & $16.7\pm0.5$ pb \\[0.1cm]
\hline
\end{tabular}
\end{table}

% End of text imported from the .//sections/inclXsection.tex input file

\subsection{Measurement of differential cross-sections}
\label{sec:DiffXsec}

% The next lines are included from the .//sections/diffXsection.tex input file
The differential cross-section is measured by subtracting the expected background from observed data events in each bin of the studied observable in the fiducial phase space, and unfolding to correct for the detector effects, including efficiency, resolution and acceptance.
Moreover, events falling outside the fiducial region might still satisfy the detector-level selection criteria because of resolution effects; the corresponding fraction are subtracted in this measurement.
All these effects can be accounted for with a response matrix, the element $M_{dp}$ of which is defined as the
probability that an event in particle-level bin $p$ is reconstructed at detector-level in bin $d$.
The diagonal elements denote the fraction of events in a detector-level bin that originate from the same bin at particle-level, which defines the unfolding purity.
 
The unfolding is performed by calculating the response matrix using the signal samples, and applying it to the distribution of the measured observable at detector-level.
To reduce the statistical uncertainty while constraining the regularisation bias, an iterative unfolding method based on Bayes’ theorem ~\cite{DAgostini:1994fjx}, with the particle-level distribution from MC as a prior, is used. The number of iterations for the Bayesian unfolding is chosen to be two for both \mfourl~and \ptfourl. The nominal response matrix, corrections and priors are all derived from simulation. The unfolding purity is globally greater than 80\%.
 
The statistical uncertainty in the unfolded distributions is estimated with 1000 MC sets generated from the observed distributions (pseudodata), using Poisson random values for each bin.
The root mean square of the deviation of the resulting unfolded spectrum of the pseudodata sets, from the actual unfolded data, is considered as the statistical uncertainty in each bin.
The uncertainty due to any inherent bias in the unfolding method and the residual mis-modelling in simulation is evaluated with a  procedure described in Ref.~\cite{malaescu2009iterative}.
The simulations are reweighted by a polynomial function to match the observed data when unfolded with the nominal response matrix. After unfolding, the differences between the reweighted MC distribution and the measured data distribution define the unfolding bias uncertainty in each bin, which remains globally below 1\%, owing to the high resolution of the measured observable. The experimental and theoretical uncertainties discussed in Section~\ref{sec:syst} are estimated separately for signal and background.
For the signal process, each uncertainty variation leads to a modification of the response matrix and therefore results in a difference in the unfolded distribution. The background uncertainties impact the data subtraction before unfolding.
 
Figure~\ref{fig:differencial_xs} presents the measured differential cross-sections for \mfourl~ and
\ptfourl. The unfolded distributions are compared with the SM predictions, accounting for production of \qqZZ, \ggZZ, and EW \qqZZjj. The EW \qqZZjj~contribution is included in the \qqZZ~ category in the figure.
The results agree well within the total statistical and systematic uncertainty in most of the bins.
For the \mfourl~distribution, the two predictions, both with and without EW corrections, describe the data adequately, while for the \ptfourl~distribution, \progname{Sherpa} gives a better description of the low-\ptfourl region, due to its merging and matching of multiple partons in the predictions. The impact of the NLO EW correction is generally not large compared to the precision of the measurements, but for MATRIX the description with the correction included is slightly better than without. In the case of \progname{Sherpa}, the predictions without NLO EW corrections are slightly better than those including them, correlated with the reduced cross-section estimate discussed in Section~\ref{sec:data_mc}.

\begin{figure}[!htbp]
\centering
\subfloat[]{\includegraphics[width = 0.48\textwidth]{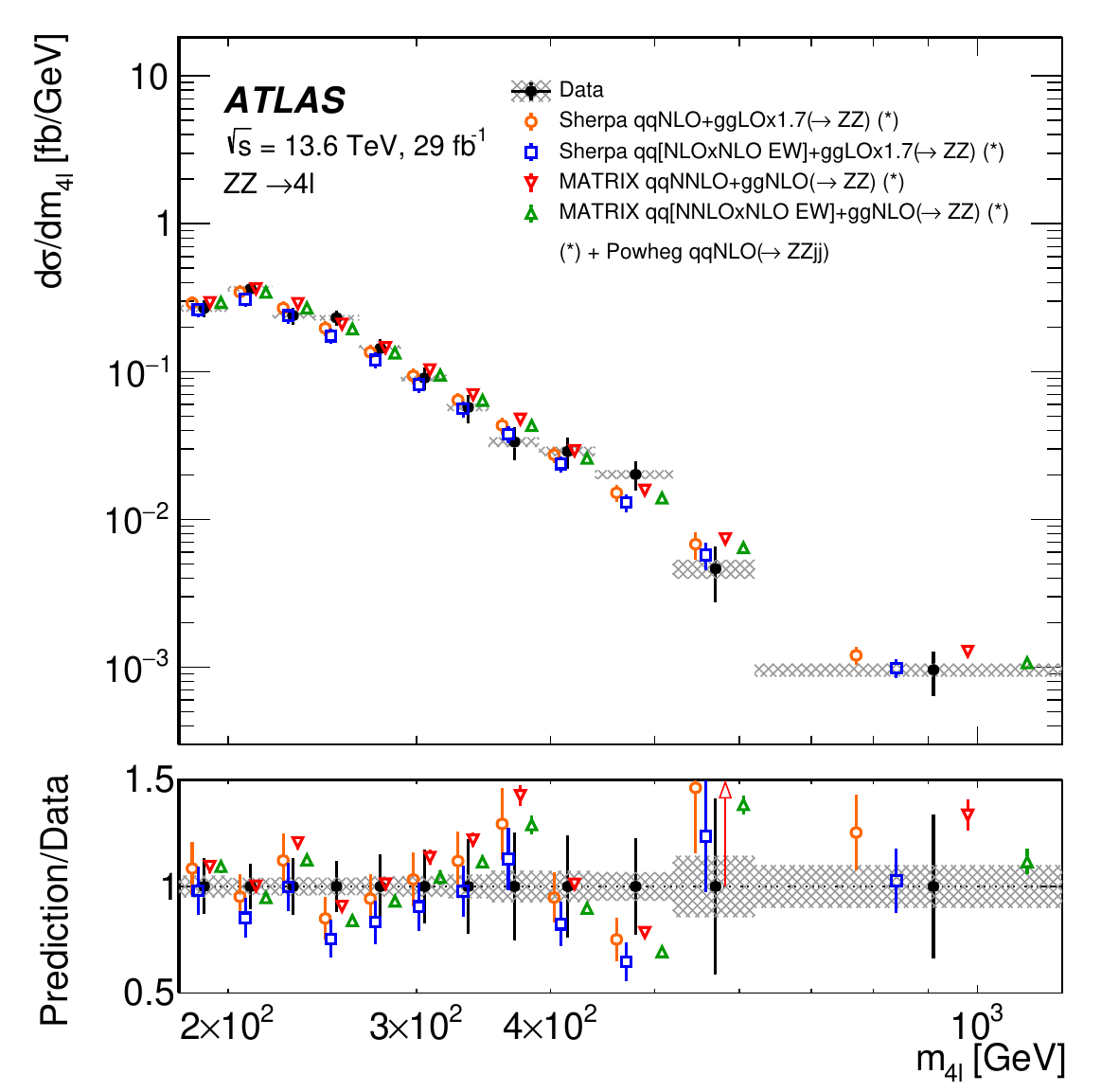}}
\subfloat[]{\includegraphics[width = 0.48\textwidth]{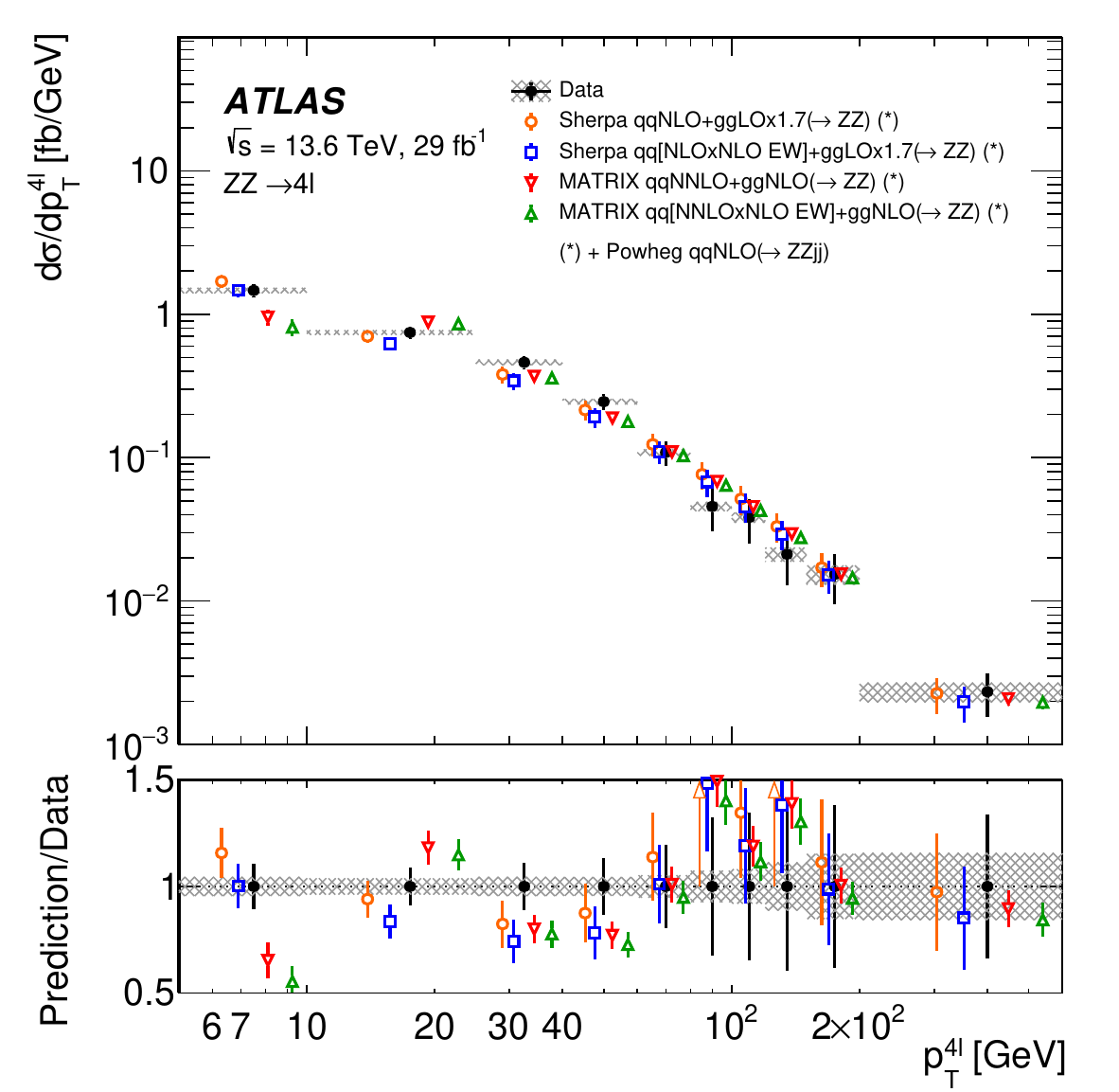}}
 
\caption{The measured differential cross-sections (filled points) are compared with the predictions, in each of the (a) \mfourl~ and (b) \ptfourl~ bins. The error bars on the measurement points give the total uncertainty and the hatched band gives the systematic uncertainty. The upwards arrows in the ratio panels indicate that the cross-section of the corresponding prediction is more than 1.5 times the data in that bin.
In all predictions, EW \qqZZjj~ is modelled by \POWHEGBOX[v2].
\progname{Sherpa} models \qqZZ ~at NLO QCD and LO EW accuracy,
and provides NLO EW corrections as alternative internal weights.
The \progname{Sherpa} \qqZZ~predictions using nominal and alternative weights are shown respectively as `$\mathrm{qqNLO}$' and `$\mathrm{qq[NLO\times NLO~EW]}$'.
The \ggZZ~ process predicted by \progname{Sherpa} is at LO QCD accuracy, and scaled to NLO QCD using a $K$-factor of 1.7.
The error bands on the \progname{Sherpa} predictions show the theoretical uncertainties in the signal process, where QCD, PDF and parton shower uncertainties are considered. MATRIX calculates \qqZZ ~at NNLO QCD, and at up to NLO EW accuracy. MATRIX calculates \ggZZ ~ at NLO QCD accuracy.
Only the QCD scale uncertainty is considered in the MATRIX predictions. The lower panels show the ratios of the SM predictions to the measured data.
\label{fig:differencial_xs}}
\end{figure}
\FloatBarrier

% End of text imported from the .//sections/diffXsection.tex input file

\section{Conclusion}

% The next lines are included from the .//sections/conclusion.tex input file
This paper presents cross-section measurements of $ZZ$ production in the \fourl~final state, based on 29~\ifb
of data collected by the ATLAS detector at the LHC in $pp$ collisions at $\sqrt{s} = 13.6~\TeV$. The inclusive cross-sections within both a fiducial and the total phase-space regions are measured. Differential measurements of \mfourl~and \ptfourl~are also presented. The results are well described by the predictions, both from state-of-the-art MC simulation and from fixed-order calculations with accuracy up to NNLO QCD + NLO EW\@.
These results are the first step in the programme of diboson production measurements at the new centre-of-mass energy of 13.6~\TeV.

% End of text imported from the .//sections/conclusion.tex input file

\section*{Acknowledgements}

% The next lines are included from the .//acknowledgements/Acknowledgements.tex input file

We thank CERN for the very successful operation of the LHC and its injectors, as well as the support staff at
CERN and at our institutions worldwide without whom ATLAS could not be operated efficiently.
 
The crucial computing support from all WLCG partners is acknowledged gratefully, in particular from CERN, the ATLAS Tier-1 facilities at TRIUMF/SFU (Canada), NDGF (Denmark, Norway, Sweden), CC-IN2P3 (France), KIT/GridKA (Germany), INFN-CNAF (Italy), NL-T1 (Netherlands), PIC (Spain), RAL (UK) and BNL (USA), the Tier-2 facilities worldwide and large non-WLCG resource providers. Major contributors of computing resources are listed in Ref.~\cite{ATL-SOFT-PUB-2023-001}.
 
We gratefully acknowledge the support of ANPCyT, Argentina; YerPhI, Armenia; ARC, Australia; BMWFW and FWF, Austria; ANAS, Azerbaijan; CNPq and FAPESP, Brazil; NSERC, NRC and CFI, Canada; CERN; ANID, Chile; CAS, MOST and NSFC, China; Minciencias, Colombia; MEYS CR, Czech Republic; DNRF and DNSRC, Denmark; IN2P3-CNRS and CEA-DRF/IRFU, France; SRNSFG, Georgia; BMBF, HGF and MPG, Germany; GSRI, Greece; RGC and Hong Kong SAR, China; ISF and Benoziyo Center, Israel; INFN, Italy; MEXT and JSPS, Japan; CNRST, Morocco; NWO, Netherlands; RCN, Norway; MEiN, Poland; FCT, Portugal; MNE/IFA, Romania; MESTD, Serbia; MSSR, Slovakia; ARIS and MVZI, Slovenia; DSI/NRF, South Africa; MICINN, Spain; SRC and Wallenberg Foundation, Sweden; SERI, SNSF and Cantons of Bern and Geneva, Switzerland; NSTC, Taipei; TENMAK, T\"urkiye; STFC/UKRI, United Kingdom; DOE and NSF, United States of America.
 
Individual groups and members have received support from BCKDF, CANARIE, CRC and DRAC, Canada; CERN-CZ, PRIMUS 21/SCI/017 and UNCE SCI/013, Czech Republic; COST, ERC, ERDF, Horizon 2020, ICSC-NextGenerationEU and Marie Sk{\l}odowska-Curie Actions, European Union; Investissements d'Avenir Labex, Investissements d'Avenir Idex and ANR, France; DFG and AvH Foundation, Germany; Herakleitos, Thales and Aristeia programmes co-financed by EU-ESF and the Greek NSRF, Greece; BSF-NSF and MINERVA, Israel; Norwegian Financial Mechanism 2014-2021, Norway; NCN and NAWA, Poland; La Caixa Banking Foundation, CERCA Programme Generalitat de Catalunya and PROMETEO and GenT Programmes Generalitat Valenciana, Spain; G\"{o}ran Gustafssons Stiftelse, Sweden; The Royal Society and Leverhulme Trust, United Kingdom.
 
In addition, individual members wish to acknowledge support from Chile: Agencia Nacional de Investigaci\'on y Desarrollo (FONDECYT 1190886, FONDECYT 1210400, FONDECYT 1230812, FONDECYT 1230987); China: National Natural Science Foundation of China (NSFC - 12175119, NSFC 12275265, NSFC-12075060); Czech Republic: PRIMUS Research Programme (PRIMUS/21/SCI/017); EU: H2020 European Research Council (ERC - 101002463); European Union: European Research Council (ERC - 948254), Horizon 2020 Framework Programme (MUCCA - CHIST-ERA-19-XAI-00), European Union, Future Artificial Intelligence Research (FAIR-NextGenerationEU PE00000013), Italian Center for High Performance Computing, Big Data and Quantum Computing (ICSC, NextGenerationEU), Marie Sklodowska-Curie Actions (EU H2020 MSC IF GRANT NO 101033496); France: Agence Nationale de la Recherche (ANR-20-CE31-0013, ANR-21-CE31-0013, ANR-21-CE31-0022, ANR-22-EDIR-0002), Investissements d'Avenir Idex (ANR-11-LABX-0012), Investissements d'Avenir Labex (ANR-11-LABX-0012); Germany: Baden-Württemberg Stiftung (BW Stiftung-Postdoc Eliteprogramme), Deutsche Forschungsgemeinschaft (DFG - 469666862, DFG - CR 312/5-1); Italy: Istituto Nazionale di Fisica Nucleare (FELLINI G.A. n. 754496, ICSC, NextGenerationEU); Japan: Japan Society for the Promotion of Science (JSPS KAKENHI JP21H05085, JSPS KAKENHI JP22H01227, JSPS KAKENHI JP22H04944, JSPS KAKENHI JP22KK0227); Netherlands: Netherlands Organisation for Scientific Research (NWO Veni 2020 - VI.Veni.202.179); Norway: Research Council of Norway (RCN-314472); Poland: Polish National Agency for Academic Exchange (PPN/PPO/2020/1/00002/U/00001), Polish National Science Centre (NCN 2021/42/E/ST2/00350, NCN OPUS nr 2022/47/B/ST2/03059, NCN UMO-2019/34/E/ST2/00393, UMO-2020/37/B/ST2/01043, UMO-2021/40/C/ST2/00187); Slovenia: Slovenian Research Agency (ARIS grant J1-3010); Spain: BBVA Foundation (LEO22-1-603), Generalitat Valenciana (Artemisa, FEDER, IDIFEDER/2018/048), La Caixa Banking Foundation (LCF/BQ/PI20/11760025), Ministry of Science and Innovation (MCIN \& NextGenEU PCI2022-135018-2, MICIN \& FEDER PID2021-125273NB, RYC2019-028510-I, RYC2020-030254-I, RYC2021-031273-I, RYC2022-038164-I), PROMETEO and GenT Programmes Generalitat Valenciana (CIDEGENT/2019/023, CIDEGENT/2019/027); Sweden: Swedish Research Council (VR 2018-00482, VR 2022-03845, VR 2022-04683, VR grant 2021-03651), Knut and Alice Wallenberg Foundation (KAW 2017.0100, KAW 2018.0157, KAW 2018.0458, KAW 2019.0447); Switzerland: Swiss National Science Foundation (SNSF - PCEFP2\_194658); United Kingdom: Leverhulme Trust (Leverhulme Trust RPG-2020-004); United States of America: U.S. Department of Energy (ECA DE-AC02-76SF00515), Neubauer Family Foundation.

% End of text imported from the .//acknowledgements/Acknowledgements.tex input file

\printbibliography
 
\clearpage
% ATLAS Collaboration author list
% Reference date of STDM-2022-17 is 2023-08-08
% Author list last updated on date 23-AUG-24
% Data extracted on 23-Aug-2024 for paper reference STDM-2022-17
% at 4:27pm
 
\begin{flushleft}
\hypersetup{urlcolor=black}
{\Large The ATLAS Collaboration}

\bigskip

\AtlasOrcid[0000-0002-6665-4934]{G.~Aad}$^\textrm{\scriptsize 102}$,
\AtlasOrcid[0000-0002-5888-2734]{B.~Abbott}$^\textrm{\scriptsize 120}$,
\AtlasOrcid[0000-0002-1002-1652]{K.~Abeling}$^\textrm{\scriptsize 55}$,
\AtlasOrcid[0000-0001-5763-2760]{N.J.~Abicht}$^\textrm{\scriptsize 49}$,
\AtlasOrcid[0000-0002-8496-9294]{S.H.~Abidi}$^\textrm{\scriptsize 29}$,
\AtlasOrcid[0000-0002-9987-2292]{A.~Aboulhorma}$^\textrm{\scriptsize 35e}$,
\AtlasOrcid[0000-0001-5329-6640]{H.~Abramowicz}$^\textrm{\scriptsize 151}$,
\AtlasOrcid[0000-0002-1599-2896]{H.~Abreu}$^\textrm{\scriptsize 150}$,
\AtlasOrcid[0000-0003-0403-3697]{Y.~Abulaiti}$^\textrm{\scriptsize 117}$,
\AtlasOrcid[0000-0002-8588-9157]{B.S.~Acharya}$^\textrm{\scriptsize 69a,69b,m}$
% ...
\bigskip
\\

% ...
$^{29}$Physics Department, Brookhaven National Laboratory, Upton NY; United States of America.\\
% ...
$^{35}$$^{(a)}$Facult\'e des Sciences Ain Chock, Universit\'e Hassan II de Casablanca;$^{(b)}$Facult\'{e} des Sciences, Universit\'{e} Ibn-Tofail, K\'{e}nitra;$^{(c)}$Facult\'e des Sciences Semlalia, Universit\'e Cadi Ayyad, LPHEA-Marrakech;$^{(d)}$LPMR, Facult\'e des Sciences, Universit\'e Mohamed Premier, Oujda;$^{(e)}$Facult\'e des sciences, Universit\'e Mohammed V, Rabat;$^{(f)}$Institute of Applied Physics, Mohammed VI Polytechnic University, Ben Guerir; Morocco.\\
% ...
$^{49}$Fakult\"{a}t Physik , Technische Universit{\"a}t Dortmund, Dortmund; Germany.\\
% ...
$^{55}$II. Physikalisches Institut, Georg-August-Universit\"{a}t G\"ottingen, G\"ottingen; Germany.\\
% ...
$^{69}$$^{(a)}$INFN Gruppo Collegato di Udine, Sezione di Trieste, Udine;$^{(b)}$ICTP, Trieste;$^{(c)}$Dipartimento Politecnico di Ingegneria e Architettura, Universit\`a di Udine, Udine; Italy.\\
% ...
$^{102}$CPPM, Aix-Marseille Universit\'e, CNRS/IN2P3, Marseille; France.\\
% ...
$^{117}$Department of Physics, New York University, New York NY; United States of America.\\
% ...
$^{120}$Homer L. Dodge Department of Physics and Astronomy, University of Oklahoma, Norman OK; United States of America.\\
% ...
$^{150}$Department of Physics, Technion, Israel Institute of Technology, Haifa; Israel.\\
$^{151}$Raymond and Beverly Sackler School of Physics and Astronomy, Tel Aviv University, Tel Aviv; Israel.\\
% ...
$^{a}$ Also Affiliated with an institute covered by a cooperation agreement with CERN.\\
$^{b}$ Also at An-Najah National University, Nablus; Palestine.\\
$^{c}$ Also at Borough of Manhattan Community College, City University of New York, New York NY; United States of America.\\
$^{d}$ Also at Center for High Energy Physics, Peking University; China.\\
$^{e}$ Also at Center for Interdisciplinary Research and Innovation (CIRI-AUTH), Thessaloniki; Greece.\\
$^{f}$ Also at Centro Studi e Ricerche Enrico Fermi; Italy.\\
% ...
$^{m}$ Also at Department of Physics, King's College London, London; United Kingdom.\\
% ...

\end{flushleft}

% Created with Glance <Atlas.Glance@cern.ch>

\end{document}